\documentclass[conference]{IEEEtran}

\usepackage{graphicx} 
\usepackage{booktabs} 
\usepackage{amsmath}
\usepackage{amsfonts}
\usepackage{amssymb}
\usepackage{epstopdf}

\usepackage{dblfloatfix}
\usepackage[caption=false,font=footnotesize]{subfig}
\usepackage{color}

\usepackage{cite}

\begin{document}

\title{Power Talk in DC Micro Grids: Constellation Design and Error Probability Performance}

\author{\IEEEauthorblockN{Marko Angjelichinoski, \v Cedomir Stefanovi\' c, Petar Popovski}
\IEEEauthorblockA{Department of Electronic Systems\\
Aalborg University\\
Email: \{maa,cs,petarp\}@es.aau.dk }
\and
\IEEEauthorblockN{Frede Blaabjerg}
\IEEEauthorblockA{Department of Energy Technology\\
Aalborg University\\
Email:fbl@et.aau.dk}}

\maketitle

\begin{abstract}
Power talk is a novel concept for communication among units in a Micro Grid (MG), where information is sent by using power electronics as modems and the common bus of the MG as a communication medium. 
The technique is implemented by modifying the droop control parameters from the primary control level.
In this paper, we consider power talk in a DC MG and introduce a channel model based on Thevenin equivalent. 
The result is a channel whose state that can be estimated by both the transmitter and the receiver.
Using this model, we present design of symbol constellations of arbitrary order and analyze the error probability performance.
Finally, we also show how to design adaptive modulation in the proposed communication framework, which leads to significant performance benefits.
\end{abstract}

\IEEEpeerreviewmaketitle

\section{Introduction}

The aspect of control is essential for Micro Grids (MG) in islanded mode of operation.
Advanced networked control systems are traditionally used to provide efficient and reliable operation of the MG \cite{ref1,ref2}.
Recent advances in MG control suggest that the operation of the system should not be critically dependent on an external communication system \cite{ref3}.
The trend motivates development of communication strategies for emerging power systems that do not require external communication network, reuse the existing power line equipment as a communication medium and thus offer communication reliability equivalent to the reliability of the power transmission system of the MG \cite{ref4,ref5}. 
 
In \cite{pt_gc}, we have introduced \textit{power talk}, a novel communication concept where the power electronic equipment is also used as communication modems. 
Power talk exploits the flexibility of the outer droop control loop of the Voltage Source Converter (VSC) units in the system that control the common bus parameters and determine the power flow.
The information is sent by modifying the droop control parameters and modulating information into subtle power deviations that can be detected by other units in the system.
The technique effectively transforms the MG into a communication channel and circumvents the necessity for an external communication network.
Moreover, power talk is implemented within a primary control loop and it only requires software modifications in the power electronic inverters.
This is a significant advantage compared to standard power line communication that requires installation of an additional hardware \cite{ref5}.

In \cite{pt_gc} power talk has been developed and tested for single-bus DC MG system, where the focus is on one-way communication using binary symbol constellations.
There, we have defined the communication channel in a way that the main channel impairment comes from the arbitrary variation of the load that results in random shifts of the bus voltage. 
In addition, the system configuration in \cite{pt_gc} is unknown to each VSC unit locally, making the channel generally unknown.
The focus in \cite{pt_gc} is on enabling reliable communication over the channel that does not require precise knowledge on neither the system configuration nor the load, but using a special input symbol in a role of a pilot, we have transformed the power talk communication channel into well-studied binary channels. 

In this paper we employ a different approach to address the problem of unknown system configuration and arbitrary load variations.
We present the system, as seen from the transmitting VSC, with the Thevenin equivalent, whose parameters (the equivalent resistance and equivalent voltage source) represent the channel state and can be estimated.
Assuming channel state information at both the transmitter and the receiver, we design general signaling constellations for power talk and analyze their performance. 
{Particularly, we show that the MG can be modeled as communication channel with state-dependent additive noise and symbol ``attenuation'', and asses its impact on the choice of constellation in terms of symbol error probability.
We also show that the symbol ``attenuation'' of the channel can be mitigated and the error probability performance significantly improved by using adaptive modulation, in which the choice of constellation adapts to the channel state.}

The rest of the paper is organized as follows.
Section II presents the general model and summarizes the relevant concepts introduced in \cite{pt_gc}.
Section III describes the basic mechanism of symbol transmission and detection in power talk.
Section IV focuses on designing power talk symbol constellations.
Section V assess the performance of some representative constellation designs in terms of symbol error probability and provides guidelines for designing adaptive modulation.
Finally, section VI concludes the paper.

\section{Model}

\begin{figure}[t]
\centering
\includegraphics[scale=0.3]{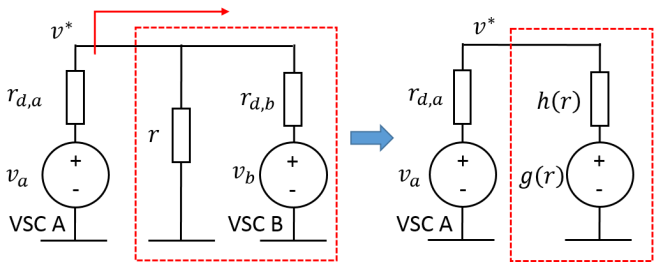}
\caption{Steady-state model of a Micro Grid and its Thevenin equivalent}
\label{DCMG_general}
\end{figure}

The steady-state model of a basic DC MG system with two VSC units that supply a load is shown on Fig.~\ref{DCMG_general}.
All units are connected to a common bus through feeders with negligible resistances.
Each VSC unit is represented with two controllable parameters: the reference voltage $v_a$ and $v_b$ and the virtual resistance $r_{d,a}$ and $r_{d,b}$. 
The load $r$ is purely resistive.

\subsection{One-Way Communication Channel}

We focus on one-way communication where VSC A transmits information to VSC B.
Thus, VSC B keeps the parameters $v_b$ and $r_{d,b}$ fixed during the transmission.
We assume that the time axis is slotted and the transmitter and the receiver are slot-synchronized.
The duration of each slot is denoted with $T_S$.
VSC A modulates information into the DC voltage level of the common bus by varying the controllable steady-state parameters $v_a$ and $r_{d,a}$.
Thus, $\mathbf{x}_a=(v_a,r_{d,a})$ is the input in the communication channel.
The transmitter inserts a specific symbol in a single slot of duration $T_S$.\footnote{Note that the duration $T_S$ in power talk is of the order of milliseconds to allow the system to reach a steady-state.}
The bus voltage is the channel output as observed by the receiver $y_b\equiv v^*=f(\mathbf{x}_a,r)+z_b$, where $f$ determines the output voltage as function of the channel input and the momentary value of the load and $z_b\sim\mathcal{N}(0,\sigma_m^2)$ is the measurement noise. 
For the model on Fig.~\ref{DCMG_general}, the steady state bus voltage is:
\begin{equation}\label{therip}
v^*=\frac{v_ar_{d,a}^{-1}+v_br_{d,b}^{-1}}{r_{d,a}^{-1}+r_{d,b}^{-1}+r^{-1}}.
\end{equation}
The receiver samples the bus voltage with frequency $f_s$ and obtains $N$ samples $y_b[n],n=1,...,N$ per symbol interval.
Note that, the transmitter also observes the channel output $y_a\equiv v^*=f(\mathbf{x}_a,r)+z_a$, where $z_a\sim\mathcal{N}(0,\sigma_m^2)$, i.e., the system provides instantaneous feedback to the transmitter.

The load $r$ changes sparsely in time. We model it as an unknown state of the channel, described with the random variable $R\sim p_R(r)$ that changes slowly w.r.t. the symbol rate.
Appropriate strategy to communicate in these settings would be to estimate $r$ using training sequences known both at the transmitter and the receiver.
This requires the transmitter to have knowledge of the input-output function $f$.
However, $f$ is also determined by the parameters of the receiver, as shown with \eqref{therip}, which are generally unknown to the transmitter.
Moreover, in general systems with multiple controllable units, the number of unknown parameters is large and obtaining detailed knowledge of $f$ is infeasible in practice.

\begin{figure}[t]
\centering
\includegraphics[scale=0.2]{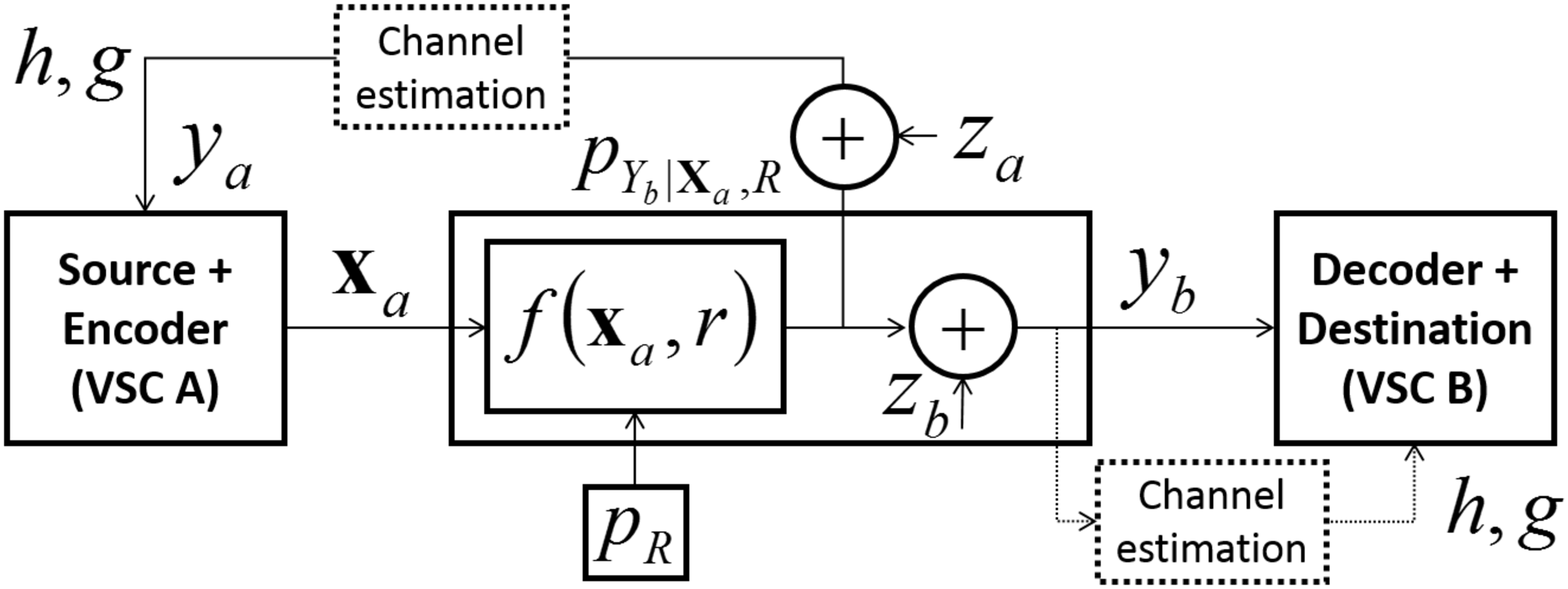}
\caption{The general one way communication channel model.}
\label{General_channel}
\end{figure}

Using Thevenin theorem, each VSC unit sees the rest of the system that is connected to the common bus as equivalent voltage source, as shown on Fig.~\ref{DCMG_general}.
The parameters of the equivalent representation are denoted with $h$ (the equivalent system resistance) and $g$ (the equivalent voltage source) and they are functions of $r$ as well as the rest of the system (which includes the parameters of the receiver).
For the simple system on Fig.~\ref{DCMG_general}, the equivalent circuit parameters are:
\begin{equation}\label{equiv_param}
g=\frac{v_br}{r+r_{d,b}},\;h=\frac{rr_{d,b}}{r+r_{d,b}}.
\end{equation}
Then, the bus voltage level as output of the communication channel can be written in the general form:
\begin{equation}\label{channel_out}
v^*=\frac{v_ar_{d,a}^{-1}+gh^{-1}}{r_{d,a}^{-1}+h^{-1}}.
\end{equation}
Once the transmitter and/or the receiver obtain estimates of $h$ and $g$, they implicitly obtain aggregate knowledge on the system and its state.
Thus, $h$ and $g$ represent the state of the communication channel and we refer to them as channel parameters, reducing the effect of the generally unknown function $f$ to two estimable parameters.
The communication model of the MG system shown on Fig.~\ref{DCMG_general}, is summarized  on Fig.~\ref{General_channel}.
The channel parameters $h$ and $g$ can be estimated by both the receiver and the transmitter using training sequence of symbols $\mathbf{x}_a$ and employing common equivalent resistance estimation technique.
Since the state of the channel changes slowly, the training sequence can be inserted periodically or whenever the transmitter detects a change of the state; recall from Fig.~\ref{General_channel} that the transmitter also observes the output from the channel.
In the remaining text we assume that both the receiver and transmitter perfectly know the instantaneous channel parameters.

\subsection{The Signaling Space}

\begin{figure}[t]
\centering
\includegraphics[scale=0.5]{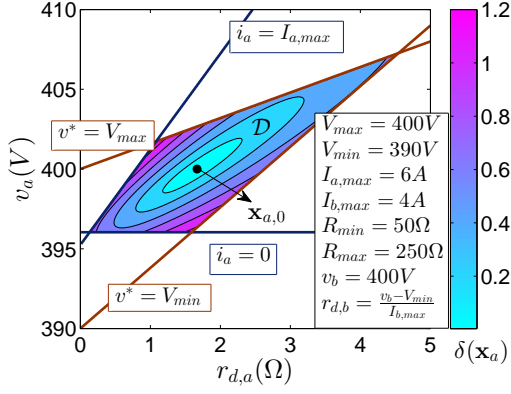}
\caption{The signaling space $\mathcal{D}$.}
\label{SSpace}
\end{figure}

The MG as a power supply system imposes certain operational limitations, such as maximum allowable bus voltage deviation and/or current deviation.
All these limitations, organized in set of constraints $\mathcal{C}$ define the signaling space $\mathcal{D}$ as the set of all input symbols $\mathbf{x}_a$ that do not violate the constraints in $\mathcal{C}$.
The signaling space for the system on Fig.~\ref{DCMG_general} is illustrated on Fig.~\ref{SSpace} for the constraint set $\mathcal{C}=\left\{V_{min}\leq v^*\leq V_{max},0\leq i_a\leq I_{a,max}\right\}$, and some representative values for the system parameters assuming that $r$ changes within some finite interval $[R_{min},R_{max}]$.
We fix the values for the system parameters and use them in the rest of the paper.
The boundaries of $\mathcal{D}$ represent the output voltage $v^*$ and current $i_a$ at $r=R_{min}$ and $r=R_{max}$, and they are easily obtained using equation \eqref{channel_out} with the corresponding channel parameters: $h(R_{min})$, $h(R_{max})$, $g(R_{min})$ and $g(R_{max})$.

\subsection{Power Deviation}

We place the power talk symbols and design signaling constellations within $\mathcal{D}$, thereby guaranteeing that the operational constraints of the MG will not be violated.
Nevertheless, each symbol $\mathbf{x}_a\in\mathcal{D}$ results in different output power supplied to the load $P(\mathbf{x}_a,r)=\frac{(v^*)^2}{r}$.
In essence, the proposed power talk communication concept deviates power in the system with each input symbol.
Let $\mathbf{x}_{a,0}=(v_{a,0},r_{d,a,0})$ denote the parameters of VSC A in normal mode of operation, i.e., when not using power talk.
We refer to this point in the signaling space as the nominal operating point or pilot, corresponding to the nominal output power $P(\mathbf{x}_{a,0},r)=\frac{(v_0^*)^2}{r}$.
We introduce the relative power deviation w.r.t. $\mathbf{x}_{a,0}$:
\begin{equation}\label{power_dev}
\delta(\mathbf{x}_{a})=\frac{\sqrt{\mathbb{E}_{R}\left\{[P(\mathbf{x}_{a},r)-P(\mathbf{x}_{a,0},r)]^2\right\}}}{\mathbb{E}_{R}\left\{P(\mathbf{x}_{a,0},r)\right\}} ,
\end{equation}
averaged {over $R$}. $\delta(\mathbf{x}_{a})$ can be thought of as a cost assigned to each input symbol $\mathbf{x}_a\in\mathcal{D}$.
Fig.~\ref{SSpace} shows $\delta$ (in \%) for symbols from $\mathcal{D}$, assuming the uniform distribution $R\sim\mathcal{U}[R_{min},R_{max}]$.
In practical systems with strict constraints that strive to minimize the power deviation introduced with power talk, one should choose symbols in the close vicinity of the pilot, making the region of small $\delta(\mathbf{x}_a)$ of practical importance.
To complete the communication model, we introduce average power deviation constraint per input symbol $\mathbf{x}_a$:
\begin{equation}\label{power_constraint}
\delta(\mathbf{x}_a)\leq\gamma.
\end{equation}
Using geometrical representation, we place the symbols around the pilot within the convex, ellipse-like hull, as shown in Fig.~\ref{SSpace}, obtained as a solution of the equation $\delta(\mathbf{x}_a)=\gamma$.

\section{Communication Principles}

In this section we illustrate the basic principles of power talk with known channel parameters.
We show that the information carrying object that can be detected in power talk is {a line (a one-dimensional affine subspace) in the signaling space}. 
Therefore, the analysis and the dimensioning of the system is conducted in an equivalent space of output lines, referred to as detection space.

\subsection{Encoding and the Detection Space}

\begin{figure}[t]
\centering
\includegraphics[scale=0.5]{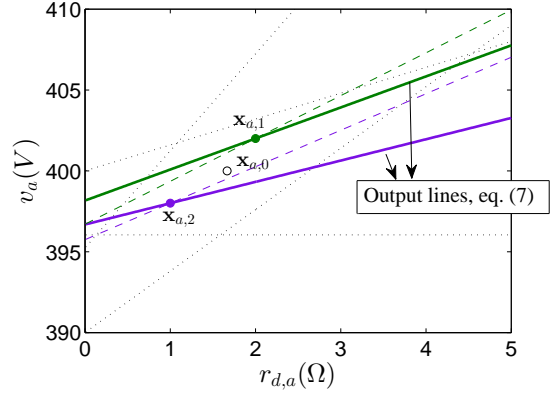}
\caption{Encoding information in lines.}
\label{SSpace_ex}
\end{figure}
\begin{figure}[t]
\centering
\includegraphics[scale=0.5]{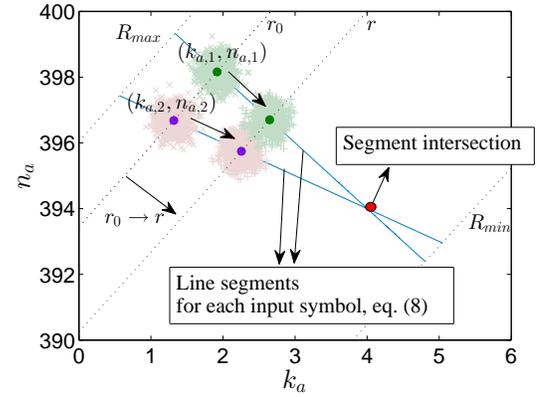}
\caption{The detection space.}
\label{DSpace_ex}
\end{figure}

\begin{figure}[t]
\centering
\includegraphics[scale=0.5]{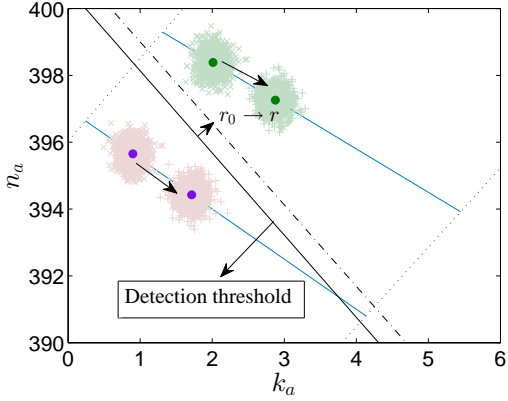}
\caption{The decision regions for binary power talk.}
\label{DSpace_noise_bin}
\end{figure}

We first assess binary signaling in order to present key insights.
The transmitter encodes the data using two symbols: `$0$' is encoded as a symbol $\mathbf{x}_{a,1}$, producing the output voltage $v_1^*$ and  `$1$' is encoded as $\mathbf{x}_{a,2}$, producing $v_2^*$.
The receiver observes a scalar output of the channel, namely $y_b=v^*+z_b$.
The output of the channel $y_b$ in terms of the input $\mathbf{x}_a$ is:
\begin{equation}\label{in_out}
\mathbf{S} \, \mathbf{x}_a=y_b,
\end{equation}
where $\mathbf{S}=[1\;-\frac{y_b-g}{h}]\in\mathbb{R}^2$.
Since the channel is a mapping $f:\mathbb{R}^2\rightarrow\mathbb{R}$, the receiver is unable to determine $\mathbf{x}_a$ explicitly because \eqref{in_out} is under-determined and there are many input symbols that produce the same output.
Rather, the receiver can detect the set of all the symbols that produce the observed output.
Considering the model \eqref{channel_out} and \eqref{in_out}, this set is a line:
\begin{equation}\label{lines}
v_a=r_{d,a}\bigg(\frac{y_b-g}{h}\bigg)+y_b=r_{d,a} k_a + n_a,
\end{equation}
where $k_a$ and $n_a$ are the line slope and intercept, respectively.
This motivates the notion of detection space as the set of $(k_a, n_a)$ pairs where demodulation and detection is performed.
Reordering \eqref{lines}, the output pair  $(k_a, n_a)$ in the detection space can be represented in terms of the input symbol $\mathbf{x}_a$:
\begin{equation}\label{lines_ds}
n_a=-r_{d,a}k_a+v_a .
\end{equation}
The principle of encoding information is illustrated via the following example.
Fig.~\ref{SSpace_ex} shows two arbitrary symbols in the signaling space $\mathbf{x}_{a,1}$, $\mathbf{x}_{a,2}$ and Fig.~\ref{DSpace_ex} shows the corresponding pairs $(k_{a,1},n_{a,1})$, $(k_{a,2},n_{a,2})$ in the detection space.
Assuming idealized scenario in which $z_b=0$, when the state changes from $r_0$ to $r$, the output line for each symbol $\mathbf{x}_a$ is rotated around $\mathbf{x}_a$ in the signaling space, see Fig.~\ref{SSpace_ex}.
In the detection space, the pair $(k_a, n_a)$ slides between $R_{min}$ and $R_{max}$ as $r$ varies, on the segment determined with \eqref{lines_ds}, see Fig.~\ref{DSpace_ex}.
Thus, choosing a specific symbol $\mathbf{x}_a$ within $\mathcal{D}$ generates a corresponding segment \eqref{lines_ds} in the detection space, on which the output $(k_a, n_a)$ lies, irrespective of the state of the channel.
The key idea is to determine the correct segment \eqref{lines_ds} in the detection space, that contains the observed output $(k_a, n_a)$.


When the channel state changes, an error event can occur if the two segments, corresponding to $\mathbf{x}_{a,1}$ and $\mathbf{x}_{a,2}$, intersect for some $r$.
Such an event means that the two symbols, for that channel state, {produce the same output and the receiver can not distinguish them.
{This is also illustrated in Fig.~\ref{DSpace_ex}.}
Other error events are related to the noise.
The effect of noise is the uncertainty ``cloud'' around the received line, as shown in Fig.~\ref{DSpace_ex}, which can lead the receiver to {decide incorrectly, in favor of different segment}.
The following subsection evaluates the error probability due to the noise for the binary signaling.
Section IV introduces generic approach to design high order symbol constellations in $\mathcal{D}$ to deal with segment intersections in the detection space.

\subsection{Maximum Likelihood Detector}

In presence of measurement noise and known channel parameters, we use Maximum Likelihood (ML) estimation to obtain estimate of the output line $(\hat{k}_a,\hat{n}_a)$.
Considering \eqref{lines}, to obtain the ML estimates $(\hat{k}_a, \hat{n}_a)$ we need the ML estimate of the bus voltage $\hat{v}^*$, which is simply the average of the obtained samples $y_b[n]$.
Then $\hat{n}_a=\hat{v}_\text{I}^*$ and $\hat{k}_a=\frac{g-\hat{v}_\text{II}^*}{h}$, 
where $\hat{v}_\text{I}^*$ and $\hat{v}_\text{II}^*$ are two independent estimates of $v^*$, which could be obtained by, e.g., in the first and second half of the signaling interval.
The conditional pdf of the output line $p ( (\hat{k}_a,\hat{n}_a)|\mathbf{x}_{a,i},r )$, given the state $r$ (i.e., the corresponding $h$ and $g$),  is:
\begin{equation}\label{line_est}
(\hat{k}_a,\hat{n}_a)\sim\mathcal{N}\bigg(\bigg(\frac{v^*-g}{h},v^*\bigg),\text{diag}\bigg(\frac{\sigma^2}{h^2},\sigma^2\bigg)\bigg) ,
\end{equation}
where $v^*$ is given with \eqref{channel_out} and {$\sigma^2=\frac{2\sigma_m^2}{N}$.}
Since the channel parameters are known, the receiver employs ML detection (MLD) to determine the transmitted line $(k_a,n_a)$ from $(\hat{k}_a,\hat{n}_a)$.
The MLD decision rule is:
\begin{equation}\label{MLD_binary}
\ln{\frac{p((\hat{k}_a,\hat{n}_a)|\mathbf{x}_{a,1},r)}{p((\hat{k}_a,\hat{n}_a)|\mathbf{x}_{a,2},r)}}  \underset{\mathbf{x}_{a,2}}{ \overset{\mathbf{x}_{a,1}} {\gtrless}} 0,
\end{equation}
or equivalently, using \eqref{line_est}:
\begin{align}\label{M_1}
\hat{n}_a \underset{\mathbf{x}_{a,2}}{ \overset{\mathbf{x}_{a,1}} {\gtrless}} -h\hat{k}_a+(v_1^*+v_2^*)-g .
\end{align}
Fig.~\ref{DSpace_noise_bin} depicts the decision regions and their modification when the state of the channel changes.
The probability of symbol error for the binary case can be calculated as follows:
\begin{align}\label{Pe_bin1}
P(e|\mathbf{x}_{a,1},r) & =\text{Pr}\left\{\hat{n}_a<-h\hat{k}_a+(v_1^*+v_2^*)-g\right\}\\\nonumber
& =1-Q\bigg(\frac{v_2^*-v_1^*}{\sigma\sqrt{2}}\bigg) , \\\label{Pe_bin2}
P(e|\mathbf{x}_{a,2},r) & =\text{Pr}\left\{\hat{n}_a>-h\hat{k}_a+(v_1^*+v_2^*)-g\right\}\\\nonumber
& = Q\bigg(\frac{v_1^*-v_2^*}{\sigma\sqrt{2}}\bigg) .
\end{align}
The average error probability, {assuming equally-likely input symbols}, is:
\begin{equation}\label{Pe_ave_bin} 
P_e=\mathbb{E}_R\left\{Q\bigg(\frac{v_1^*-v_2^*}{\sigma\sqrt{2}}\bigg)\right\} .
\end{equation}

\begin{figure}[t]
\centering
\includegraphics[scale=0.5]{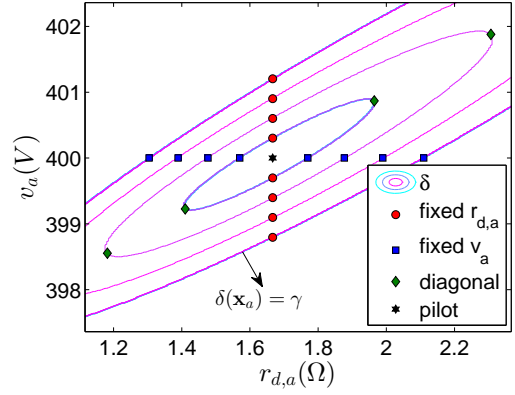}
\caption{Symbol constellations examples.}
\label{SSpace_3conts}
\end{figure}

\section{Designing Power Talk constellations}

Here, we focus on designing power talk symbol constellations of an arbitrary order.
The structure, the placement and the organization of the line segments for each symbol $\mathbf{x}_a=(v_a,r_{d,a})\in\mathcal{D}$ in the detection space (Fig.~\ref{DSpace_ex}), determines the detection regions and impacts the performance of the symbol constellation in terms of error probability.
As a general rule, good symbol constellations should produce detection space with non-intersecting segments, which are as separated from each other as possible, given the power deviation constraint $\gamma$.
 
\subsection{Channel Characterization over Different Constellations}

\begin{figure*}[t]
\centering
\subfloat[fixed $v_a$]{\includegraphics[scale=0.3]{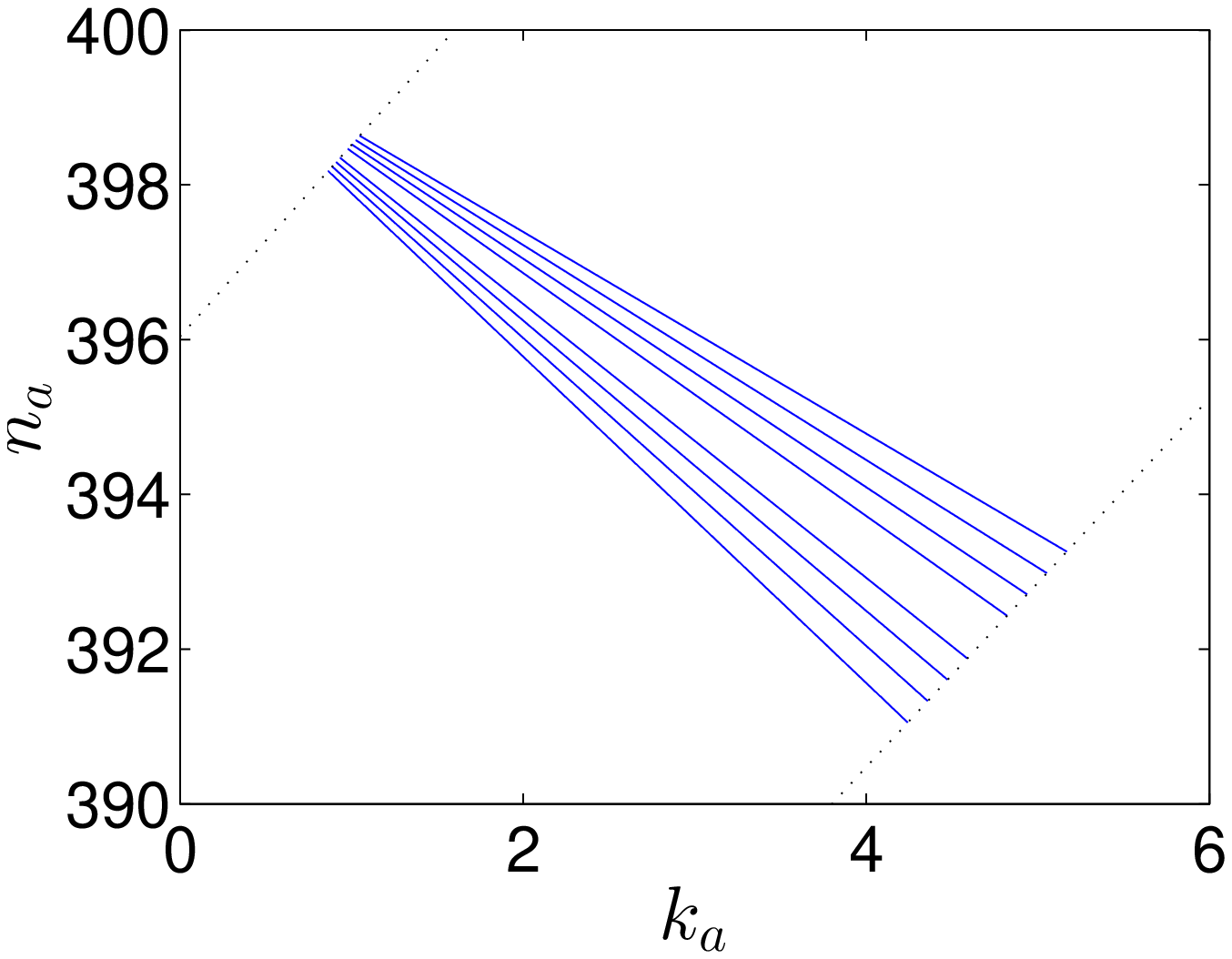} \label{DSpace_conts1}}
\hfil
\subfloat[fixed $r_{d,a}$]{\includegraphics[scale=0.3]{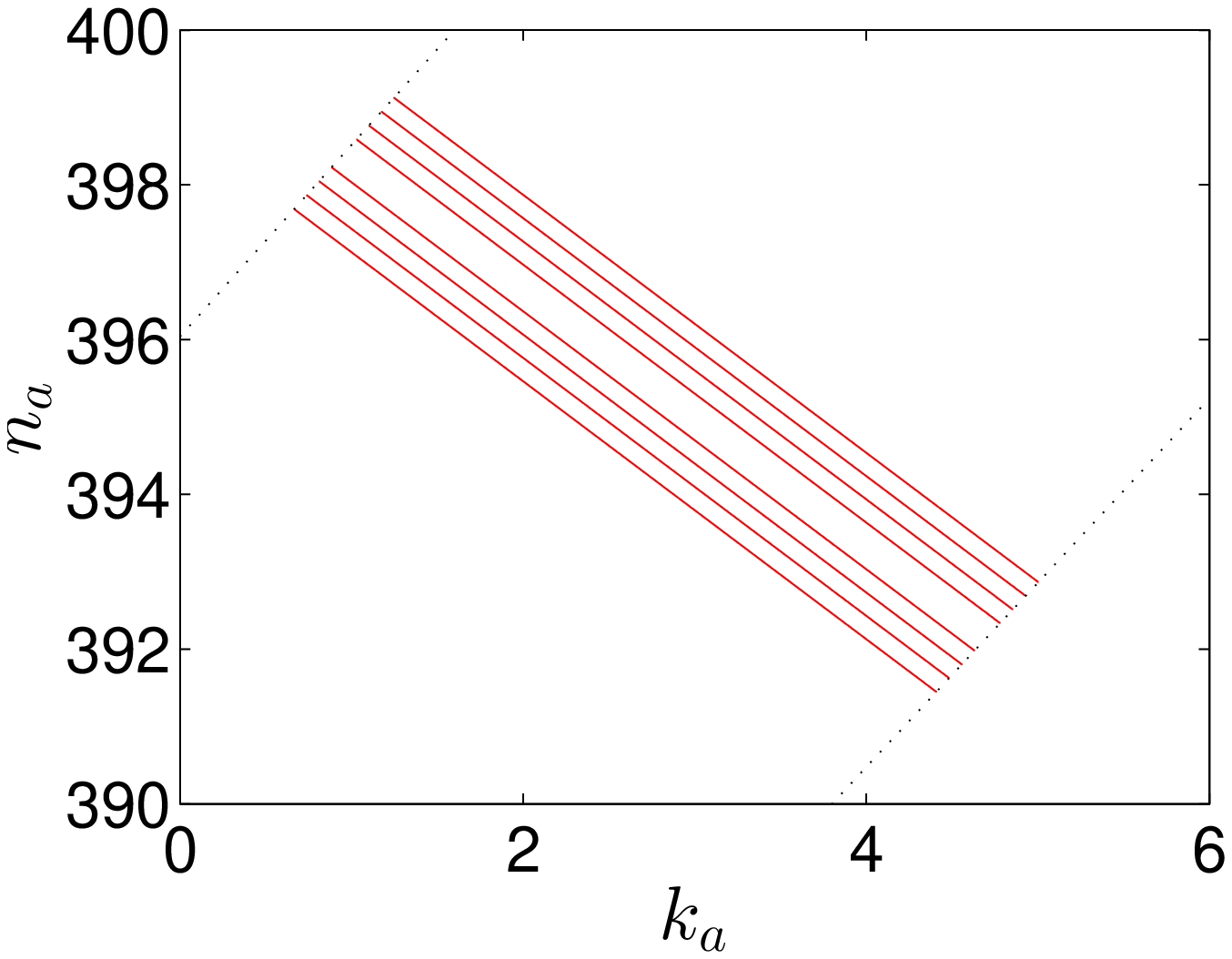} \label{DSpace_conts2}}
\hfil
\subfloat[Diagonal constellation]{\includegraphics[scale=0.3]{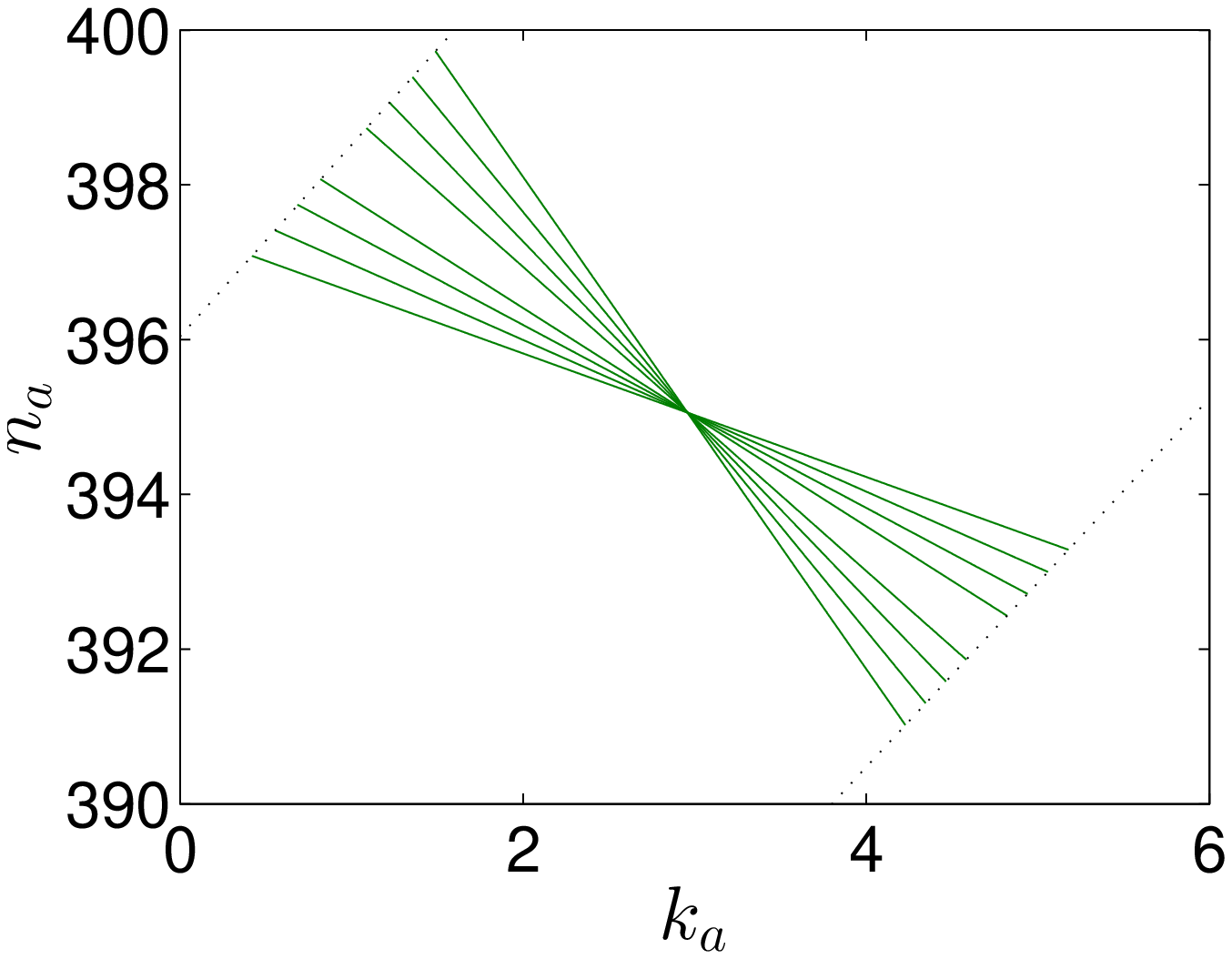} \label{DSpace_conts3}}
\caption{The symbol constellations in the detection space.}
\label{DSpace_conts}
\end{figure*}

Fig.~\ref{SSpace_3conts} depicts three examples of symbol constellations in the signaling space}: 1) fixed $v_a$ constellation, i.e., only $r_{d,a}$ is varied and $v_a=v_{a,0}$; 2) fixed $r_{d,a}$ constellation, i.e., only $v_a$ is varied while $r_{d,a}=r_{d,a,0}$ and 3) diagonal constellation, where $v_a$ is chosen to minimize the average power deviation, i.e., $v_a=\min_{v_a}\delta(\mathbf{x}_a)$, for a given $r_{d,a}$.
The number of symbols for each constellation is $M=8$.

The output of each symbol $\mathbf{x}_{a,i}$, $v_i^*$ is given with \eqref{channel_out} and the parameters of the received line are $(k_{a,i},n_{a,i})$.
The constellations are designed with the same power deviation constraint.
Each symbol $\mathbf{x}_{a,i},1<i<M$, for each constellation satisfies $\delta(\mathbf{x}_{a,i})=\gamma_i\leq\gamma$, with equality for $i=1$ and $i=M$.
The corresponding detection space for each constellation from Fig.~\ref{SSpace_3conts} is presented on Fig.~\ref{DSpace_conts}.
The fixed $v_a$ and fixed $r_{d,a}$ constellations produce detection space with non-intersecting segments over the range $r\in[R_{min},R_{max}]$.
Thus, these two constellations can achieve strictly positive information rate in any channel state.
Oppositely, the diagonal constellation leads to intersection of all segments for a single value of $r$.
In particular, for this value of the load, the output line of each symbol becomes identical to the output line produced by the pilot $\mathbf{x}_{a,0}$, i.e. $(k_{a,0},n_{a,0})$.
Thus, $P(\mathbf{x}_{a},r)=P(\mathbf{x}_{a,0},r)$ and $\delta(\mathbf{x}_a)=0$.
In other words, in this state, the channel completely attenuates each input symbol and the rate is zero.

\begin{figure}[t]
\centering
\includegraphics[scale=0.5]{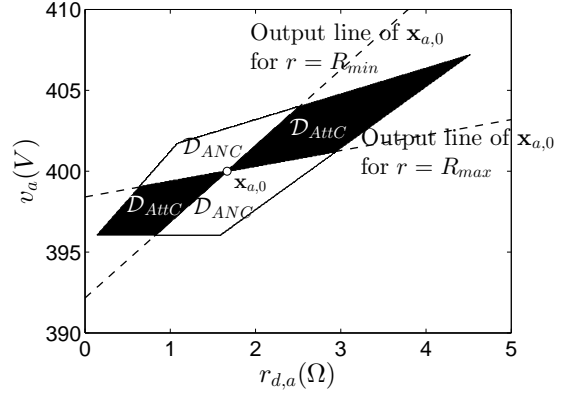}
\caption{Channel characterization on different parts of $\mathcal{D}$.}
\label{SignalingRegions}
\end{figure}

This effect is illustrated and generalized on Fig.~\ref{SignalingRegions}.
The bordering lines between the shaded (denoted with $\mathcal{D}_{AttC}$) and the bright (denoted with $\mathcal{D}_{ANC}$) regions of the signaling space are the output lines of $\mathbf{x}_{a,0}$ for $r=R_{min}$ and $r=R_{max}$.
For a constellation in $\mathcal{D}_{AttC}$, there is a corresponding state for which the channel produces identical output for each input symbol, $v_i^*=v_0^*$, and the information rate is zero. 
On the other hand, for each state in $\mathcal{D}_{ANC}$, the MG behaves like a channel with state-dependent additive noise and the achievable rate is non-negative.

\subsection{MLD and Symbol Error Probability}

The order of the constellation is denoted with $M$.
Then the MLD detector decides in favor of symbol $\mathbf{x}_{a,i}$ if:
\begin{equation}\label{MLD}
\ln{\frac{p((\hat{k}_a,\hat{n}_a)|\mathbf{x}_{a,i},r)}{p((\hat{k}_a,\hat{n}_a)|\mathbf{x}_{a,j},r)}}\geq 0,\;j\neq i .
\end{equation}
The set of all received lines that satisfy (\ref{MLD}) form the decision region for the symbol $\mathbf{x}_{a,i}$, which is denoted with $\Lambda_{i}(r)$.
Ordering the input symbols $\mathbf{x}_{a,i}$ for the index $i=1,...,M$, such that $v_i^*>v_{i+1}^*$, and using \eqref{line_est} in \eqref{MLD}, the decision region of $\mathbf{x}_{a,i}$ for $1<i<M$ is:
\begin{align}\label{reg_gen}
\Lambda_{i}(r) :\; \left\{
  \begin{array}{lr}
    \hat{n}_a<-h\hat{k}_a+(v_i^*+v_{i-1}^*)-g, \\
    \hat{n}_a>-h\hat{k}_a+(v_i^*+v_{i+1}^*)-g.
  \end{array}
\right.
\end{align}
and for $\mathbf{x}_{a,1}$ or $\mathbf{x}_{a,M}$ is:
\begin{align}\label{M_1}
 \Lambda_{1}(r) & : \;\hat{n}_a>-h\hat{k}_a+(v_1^*+v_2^*)-g, \\\label{M_M}
 \Lambda_{M}(r) & :\;\hat{n}_a<-h\hat{k}_a+(v_{M}^*+v_{M-1}^*)-g .
\end{align}
Given the state of the channel, the probability of error for the symbol $\mathbf{x}_{a,i}$ is:
\begin{align}\label{pe_per_sym}
P(\epsilon|\mathbf{x}_{a,i},r)=\int_{(\hat{n}_{a},\hat{k}_{a})\notin\Lambda_{i}(r)}p((\hat{n}_{a},\hat{k}_{a})|\mathbf{x}_{a,i},r)d\hat{n}_{a}d\hat{k}_{a}
\end{align}
Via a standard analysis it can be shown that the symbol error probability, averaged over {equally-likely} symbols and $r$, is:
\begin{equation}\label{Pe_all}
P_e=\mathbb{E}_R\left\{P(\epsilon|r)\right\}=\mathbb{E}_R\left\{\frac{2}{M}\sum_{i=2}^MQ\bigg(\frac{v_{i-1}^*-v_{i}^*}{\sigma\sqrt{2}}\bigg)\right\}.
\end{equation}

\subsection{Adaptive Modulation}

\begin{figure}[t]
\centering
\includegraphics[scale=0.5]{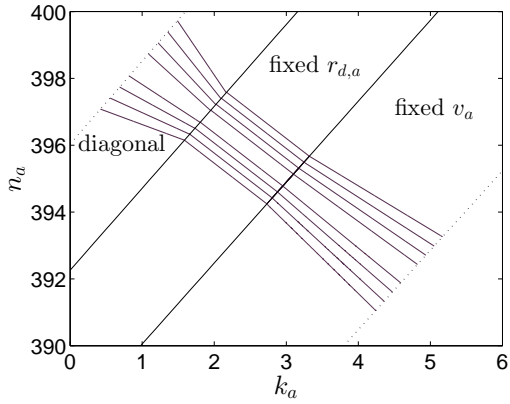}
\caption{The decision space with adaptive modulation.}
\label{Adaptive}
\end{figure}

Assuming knowledge of the channel coefficients $h$ and $g$ at the transmitter, motivates the design of an adaptive modulation. 
As the channel state changes from $R_{min}$ to $R_{max}$, the symbols constellation can be adapted in a manner that reduces the error probability and achieves positive rate, regardless of the channel state.
Consider as an example the three symbol constellations depicted on Fig.~\ref{DSpace_conts}.
The composite constellation for the adaptive modulation, made by using these three constellations, is depicted on Fig.~\ref{Adaptive}.
In the case of adaptive modulation, in each state, the MG behaves like channel with additive noise and power talk achieves non-negative rate.
It can be shown that this composite constellation minimizes the probability of error w.r.t. the given constituent constellations.
The algorithm to determine the adaptation thresholds for $h$ and $g$ (omitted here due to space limitations) only requires that the transmitter and receiver have knowledge of the constituent constellations.

\section{Evaluation}

In this section, we evaluate the power talk constellations and the adaptive modulation strategy in terms of the error probability.
We fix the standard deviation of the noise at $\sigma=0.1V$ and $\gamma=0.004$, and investigate the increase in the error probability with the order of the constellation.
We assume that the state, i.e., the load $r$, is distributed according to an uninformative prior $R\sim\mathcal{U}[R_{min},R_{max}]$.

\begin{figure}[t]
\centering
\includegraphics[scale=0.5]{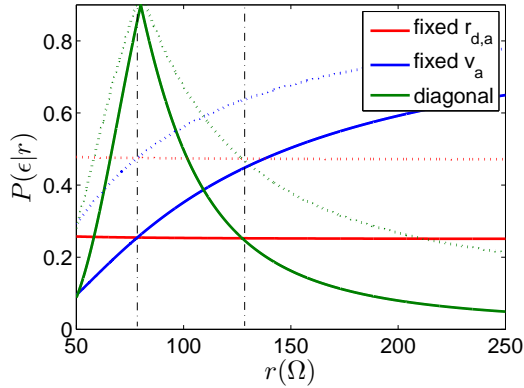}
\caption{The probability of symbol error given $r$: solid line $M=10$, dot line $M=16$.}
\label{PeR10}
\end{figure}

Fig.~\ref{PeR10} shows the error probability given the channel state $r$, $P(\epsilon|r)$ for each of the three constellations.
$P(\epsilon|r)$ is depicted over the support of $r$ to illustrate the behavior of different constellations as the state of the channel varies. 
For different channel state, each constellation performs differently in terms of error probability, with each constellation dominating the performance in different states.
The points of intersection of the probability curves for fixed $v_a$ and fixed $r_{d,a}$, as well as fixed $r_{d,a}$ and the diagonal constellations, determine the adaptation thresholds.

\begin{figure}[t]
\centering
\includegraphics[scale=0.5]{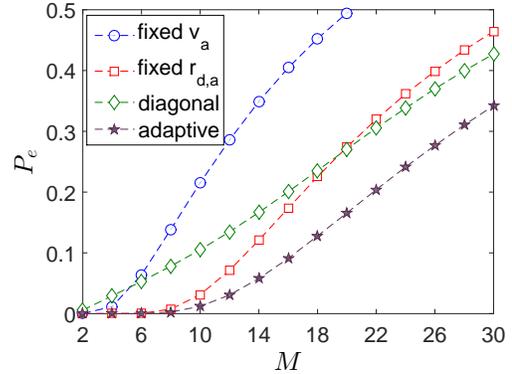}
\caption{The error probability vs. the order of the modulation.}
\label{results}
\end{figure}

Fig.~\ref{results} shows the error probability averaged over $r$, i.e. it depicts $P_e$ calculated via \eqref{Pe_all}.
Evidently, out of the three investigated constellations, the fixed $v_a$ results in the worst performance.
Using adaptive modulation across the detection space provides uniformly improved performance in terms of the error probability.
Fig.~\ref{results} suggest that if we increase the number of constellations that we use for adaptation, the error probability will decrease even further.
The derivation of the lower bound on the probability of error for adaptive modulation is left for future work.
Note that the value of the noise variance is only illustrative and in practice it is usually even lower.
This will not change the general trend of the error probability curves; however, it will allow us to pack constellations of very high order for the same power deviation constraint, providing for high symbol rates.

\section{Conclusion}

{In this paper we investigated the design of arbitrary order symbol constellation for power talk in DC microgrids  \cite{pt_gc}, assuming that the channel state information is known.
Building on the premisses of the assessed framework, we represented the problem of the communication system design and analysis in the corresponding detection space.
We characterized symbol-error performance and showed that for some symbol constellations the achievable information-rate is non-negative, irrespective of the channel state, and that there are also constellations for which the achievable information rate is zero in certain channel states. 
We also provided guidelines for designing an adaptive modulation strategy, with the aim of achieving a favorable overall symbol-error performance.}


\section*{Acknowledgment}
The work presented in this paper was supported in part by EU, under grant agreement no. 607774 ``ADVANTAGE''.

\bibliographystyle{IEEEtran}

\end{document}